\documentclass[final,3p,times,twocolumn]{elsarticle}
\usepackage[utf8]{inputenc}
\usepackage{float}
\journal{Proceedings of the European Combustion Meeting 2021}
\usepackage{etoolbox}
\usepackage{subfig}
\usepackage{hyperref}
\hypersetup{hidelinks}

\usepackage{dirtytalk}
\usepackage{titlesec}
\titlespacing\section{0pt}{12pt plus 4pt minus 2pt}{0pt plus 2pt minus 2pt}

\usepackage{dblfloatfix}    % To enable figures at the bottom of page
\biboptions{sort&compress}
\usepackage{siunitx}
\usepackage{xcolor}
\makeatletter
\patchcmd{\ps@pprintTitle}% <cmd>
  {Preprint submitted to}% <search>
  {}% <replace>
  {}{}% <succes><failure>
\makeatother
\usepackage{xpatch}
\xpatchcmd{\MaketitleBox}{\hrule}{}{}{}% remove first horizontal rule (above abstract)
\xpatchcmd{\MaketitleBox}{\hrule}{}{}{}% remoce second horizonral rule (below keywords)
\let\today\relax
\patchcmd{\emailauthor}{(#2)}{}{}{}
\patchcmd{\urlauthor}{(#2)}{}{}{}
\begin{document}
		%	\linenumbers
	\begin{frontmatter}
		\title{A load balanced chemistry model with analytical Jacobian for faster reactive simulations in OpenFOAM}
		
		\author[label1]{Bulut Tekgül\corref{cor1}}
		\address[label1]{Department of Mechanical Engineering, Aalto University School of Engineering, Puumiehenkuja 5 02150 Espoo, Finland}
		
		\address[label2]{Wärtsilä Finland Oy, Vaasa FI-65101, Finland}
		
		\cortext[cor1]{Corresponding author:}
		
		\ead{bulut.tekgul@aalto.fi}
		
		\author[label2]{Heikki Kahila}

        \author[label1]{Petteri Peltonen}	
        \author[label1]{Mahmoud Gadalla}

		\author[label1]{Ossi Kaario}
	    \author[label1]{Ville Vuorinen}
		\begin{abstract}
		In this study, we introduce a novel open-source chemistry model for OpenFOAM to speed-up the reactive computational fluid dynamics (CFD) simulations using finite-rate chemistry. First, a dynamic load balancing model called \texttt{DLBFoam} is introduced to balance the chemistry load during runtime in parallel simulations. In addition, the solution of the cell-based chemistry problem is improved by utilizing an analytical Jacobian using an open-source library called \texttt{pyJac} and an efficient linear algebra library \texttt{LAPACK}. Combination of the aforementioned efforts yields a speed-up factor 200 for a high-fidelity large-eddy simulation spray combustion case compared to the standard OpenFOAM implementation. It is worth noting that the present implementation does not compromise the solution accuracy. 
		\end{abstract}
	\end{frontmatter}

\section*{Introduction}
\label{section:introduction}
Fast, robust, and accurate reacting flow simulations are paramount for predictive modeling of new and clean combustion concepts or for improving the existing ones \cite{Poinsot2001}. A straightforward way of modeling reactive flows is through computational fluid dynamics (CFD) coupled with direct integration of the chemical kinetics, using complex finite-rate chemistry. In this approach, the thermochemical state of each computational cell is represented with a composition vector  $\phi$, which consists of pressure, temperature, and species concentrations values of the cell. The time evolution of the thermochemical state forms a system of ordinary differential equations (ODE) rising from chemical kinetics $f = \partial \phi /\partial t$. Although the direct integration approach often presents a detailed view on combustion, the computational cost of it is high due to the stiffness of the ODE system. With complex fuels, the computational time spent on chemistry often exceeds the cost of the flow solution by orders of magnitude \cite{Peters2000}. 

High computational cost of finite-rate chemistry stems from the non-linear and stiff nature of chemical kinetics. In order to ensure stability and accuracy, implicit ODE solvers (integrators) are preferred, leading to a need to describe the system Jacobian. Conventionally, ODE solvers utilize finite-difference evaluation for the Jacobian elements, which is a costly operation. In the context of finite-rate chemistry, it is often shown that an analytically evaluated Jacobian yields considerable speed-up for ODE system solution \cite{McNenly2015,Perini2012,Niemeyer2017}. In addition, the overall ODE solver performance depends on the efficiency of the underlying linear algebra routines for matrix operations \cite{Hairer1996,Imren2016,hindmarsh2005sundials}. 

Considering that most modern reacting flow solvers are operated in parallel using multiple processes, it leads to a secondary and less addressed problem in terms of using finite-rate chemistry. The computational cost for solving the system of ODEs depends on the system stiffness, which is to some extent influenced by the local thermochemical state of $\phi$. This constitutes a challenge especially in parallel simulations, where one process may become a bottleneck and thereby lead to an inherent load imbalance in multiprocessing environments. 

This paper describes a novel reacting flow solver which addresses the aforementioned three major aspects in chemistry computation 1) use of analytical system Jacobian, 2) efficient linear algebra, and 3) mitigating computational load imbalance. In particular, our implementation takes benefit of the recently published open-source software by Niemeyer et al.~\cite{Niemeyer2017} called \texttt{pyJac}, that generates C subroutines for evaluation of an analytical Jacobian for chemical kinetics systems. In addition, we have replaced the matrix operations in OpenFOAM's ODE solver with \texttt{LAPACK} \cite{lapack99} routines. Lastly, these two primary enhancements are combined together with our recently published dynamic load balancing solver library called \texttt{DLBFoam} \cite{DLBFOAM}, that redistributes the chemistry load during runtime between processors. Although there are several studies in literature that introduces similar attempts for dynamic load balancing in reacting flows \cite{Antonelli2011,Shi2012,Kodavasal2016,Muela2019}, to our knowledge \texttt{DLBFoam} is the first open-source chemistry load balancer, implemented also on an open-source software.

Although extensive improvements have been implemented especially on the \texttt{DLBFoam} side, it is still important to note that the current solver framework was first introduced as an in-house solver by Kahila et al. \cite{Kahila2019} and so far utilized in a number of studies \cite{Kahila2019,Kahila2019a,Tekgul2020,Tekgul2021,Karimkashi2020}. The following sections give a brief description of the model implementation and present benchmark results from reactive flow simulations.

 \begin{figure*}[!t]
\centering
\subfloat[Standard]{{\includegraphics[width=60mm]{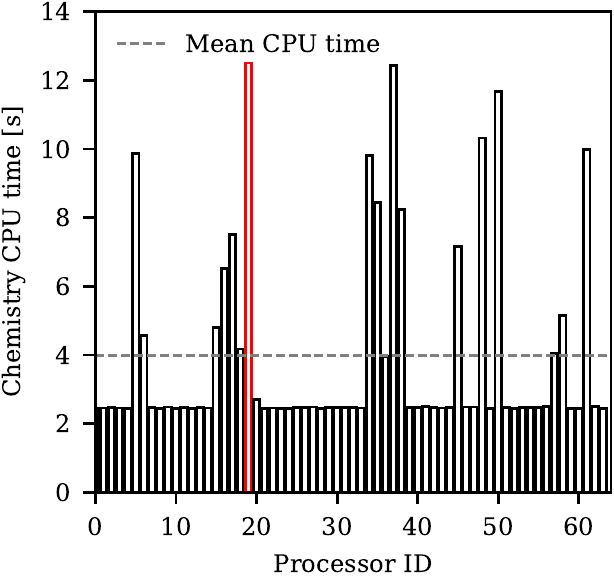} }}
\hspace{10mm}
\subfloat[DLBFoam]{{\includegraphics[width=60mm]{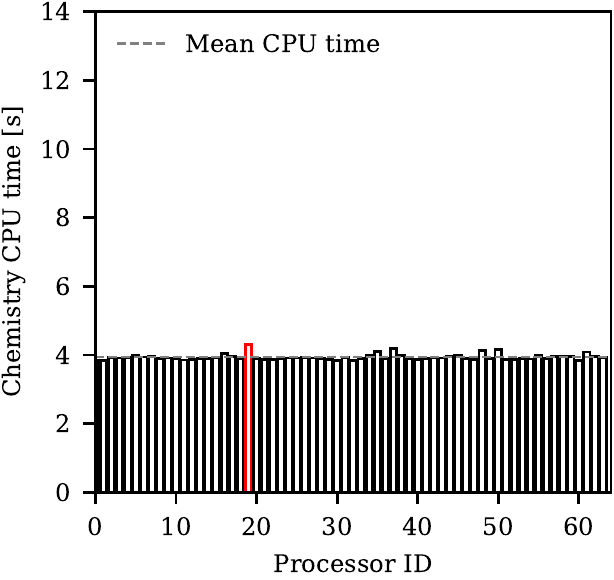} }}%
\caption{A bar plot showing the effect of dynamic load balancing when \texttt{DLBFoam} is utilized. For each case, the process with the highest computational load is marked with red. For standard OpenFOAM chemistry model (a) this process deviates greatly from the mean CPU time and creates a bottleneck. For \texttt{DLBFoam} (b), the load of the busiest process is greatly reduced via dynamic load balancing, and the bottleneck is mitigated. Data, plotting scripts, and the figure file are available under CC-BY \cite{paperdata}.}
\label{fig:imbalance}
\end{figure*}

	\section*{Methodology}
	\label{section:methods}
	\subsection*{DLBFoam: Dynamic load balancing}
	As described earlier, in finite-rate chemistry approach, the chemistry is represented as a stiff system of ODEs in each computational cell. Due to the non-linear nature of this ODE system, the computational cost of the solution may vary depending on the local thermochemical state of the problem. In parallel simulations, this behavior affects the solver performance by creating a bottleneck in the process with the highest computational load in each CFD iteration where the chemistry is solved.

    We have implemented an open-source solver library named \texttt{DLBFoam} \cite{DLBFOAM} for OpenFOAM, which utilizes dynamic load balancing to reduce the imbalance between the processes in reactive flow simulations. \texttt{DLBFoam} employs Message Passing Interface (MPI) protocol and redistributes the chemistry problems between processes, by taking chemistry problems from busy processes, and sending them to more idle ones to be solved in runtime. The solutions of the guest problems in the idle processes are then sent back to their original processes. Figure \ref{fig:imbalance} illustrates the balancing operation performed using \texttt{DLBFoam} and the difference it creates compared with OpenFOAM's standard chemistry model. It can be seen that when \texttt{DLBFoam} is used, the process with the highest computational load has a value much closer to the mean CPU time of the processes. This ensures that the busiest process does not create a large bottleneck and provides computational speed-up.

    In addition to the novel dynamic load balancing feature, \texttt{DLBFoam} also introduces a zonal reference mapping method, which further reduces the computational cost of the chemistry solution. Using Bilger's definition of the mixture fraction ($Z$) \cite{Bilger1990}, the chemistry problem in a computational cell is either solved explicitly, or mapped from a reference solution from the computational domain with a similar $Z$ value.

    In large-scale reactive CFD cases \texttt{DLBFoam} provides a speed-up around a factor of 10 compared with standard OpenFOAM chemistry solver without compromising the solution accuracy. A more thorough description of \texttt{DLBFoam} and its implementation along with benchmarking results are presented in our recent study \cite{DLBFOAM}.

	\subsection*{Analytical Jacobian and improved ODE solution}
    In addition to mitigating the computational imbalance between the processes through dynamic load balancing, we also aim for a further speedup by reducing the computing time of the chemistry solution in each computational cell. First, the open-source library \texttt{pyJac}~\cite{Niemeyer2017} is utilized to generate a fully analytical Jacobian matrix $\mathcal{J}$ with elements $\mathcal{J}_{i,j}=\partial f_i / \partial \phi_j$, to be used during the integration of the stiff ODE system for a given chemical kinetics mechanism. \texttt{pyJac} has been shown to provide much faster, accurate, and robust evaluation of the $\mathcal{J}$ compared with finite-differencing methods or other analytical evaluation approaches \cite{Niemeyer2017}. Furthermore, we have replaced all reaction rate evaluations (obtaining the right hand side of the ODE system) to corresponding function calls from \texttt{pyJac}. Second, the numerical linear algebraic operations of OpenFOAM, namely the LU decomposition and back substitution, that correspond to the ODE system integration are replaced by the more robust counterparts in the open-source library \texttt{LAPACK}~\cite{lapack99}. 
    
	\begin{figure}[!h]
    \centering
    \includegraphics[width=67mm]{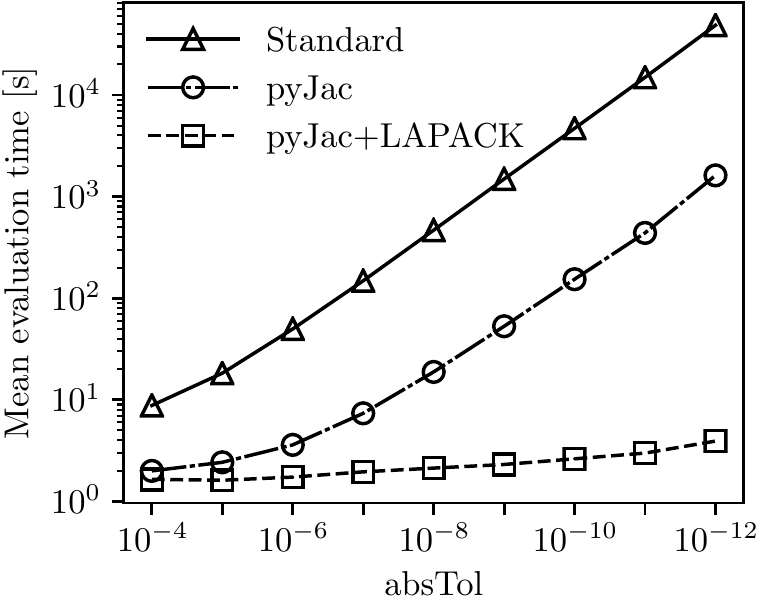}
    \caption{Mean evaluation time of a single cell chemistry problem corresponding to methane oxidation. Using \texttt{pyJac} and \texttt{LAPACK} routines keeps the evaluation time within the same magnitude order even with tighter tolerances. Data, plotting scripts, and the figure file are available under CC-BY \cite{paperdata}.}
    \label{fig:0d}
    \end{figure}
    
    Utilization of an analytical Jacobian via \texttt{pyJac} together with the robust linear algebraic routines provided by \texttt{LAPACK} allow for a speedup reaching several orders of magnitude compared with standard OpenFOAM implementation, depending on the ODE solver tolerances set for solving the chemistry problem. In \autoref{fig:0d}, a single cell methane oxidation problem is solved using the GRI-$30$ \cite{gri} kinetic mechanism with varied absolute tolerances. The advantage of the present framework is particularly demonstrated when limiting the numerical solution to tighter tolerances. There, it is shown that only the Jacobian retrieval using \texttt{pyJac} allows for about one magnitude order speedup, in terms of mean evaluation time, compared with standard routines. Moreover, handling the dense analytical Jacobian through \texttt{LAPACK} routines are capable of achieving one or more magnitude orders compared with the standard OpenFOAM routines.

	\section*{Results and discussion}
    In this section, the benchmarking results showing the performance of our novel chemistry model are introduced. A 2D shear layer and a 3D spray combustion benchmark cases are demonstrated. A 54 species, 269 reactions chemical mechanism developed by Yao et al. \cite{Yao2017} is used in both cases. For the sake of brevity, we use the name \say{\texttt{pyJac}} when we refer to the overall ODE improvements described in the previous section. However, \say{\texttt{pyJac}} name implies that both \texttt{pyJac} and the \texttt{LAPACK} routines are employed together to optimize the ODE solution procedure. 
    
\begin{figure}[!h]
	    \centering
	    \includegraphics[width=75mm]{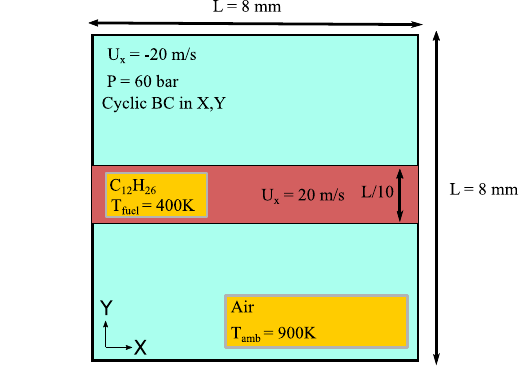}
	    \caption{A schematic describing the 2D shear layer test case configuration.}
	\label{fig:schematic}
    \end{figure}

    First, a 2D shear layer problem we introduced in the \texttt{DLBFoam} paper \cite{DLBFOAM} is investigated. This case features a 2D square domain with \SI{8}{\milli\meter} $\times$ \SI{8}{\milli\meter} dimensions and a hyperbolic tangent function to create a smooth boundary between the fuel and the oxidizer. A schematic describing the case setup is presented in Figure \ref{fig:schematic}. The domain has 400$\times$400 grid points and it is decomposed into 32 processors using Scotch decomposition algorithm \cite{scotch}. The absolute and relative ODE solver tolerances of  1e$^{-8}$ and  1e$^{-5}$ are used, respectively. The simulation is run for 1000 CFD iterations using 1) Standard OpenFOAM, 2) \texttt{DLBFoam} and 3) \texttt{DLBFoam}+\texttt{pyJac}, and the results are presented below. 	

    	\begin{figure}[H]
	    \centering
	    \includegraphics[width=64mm]{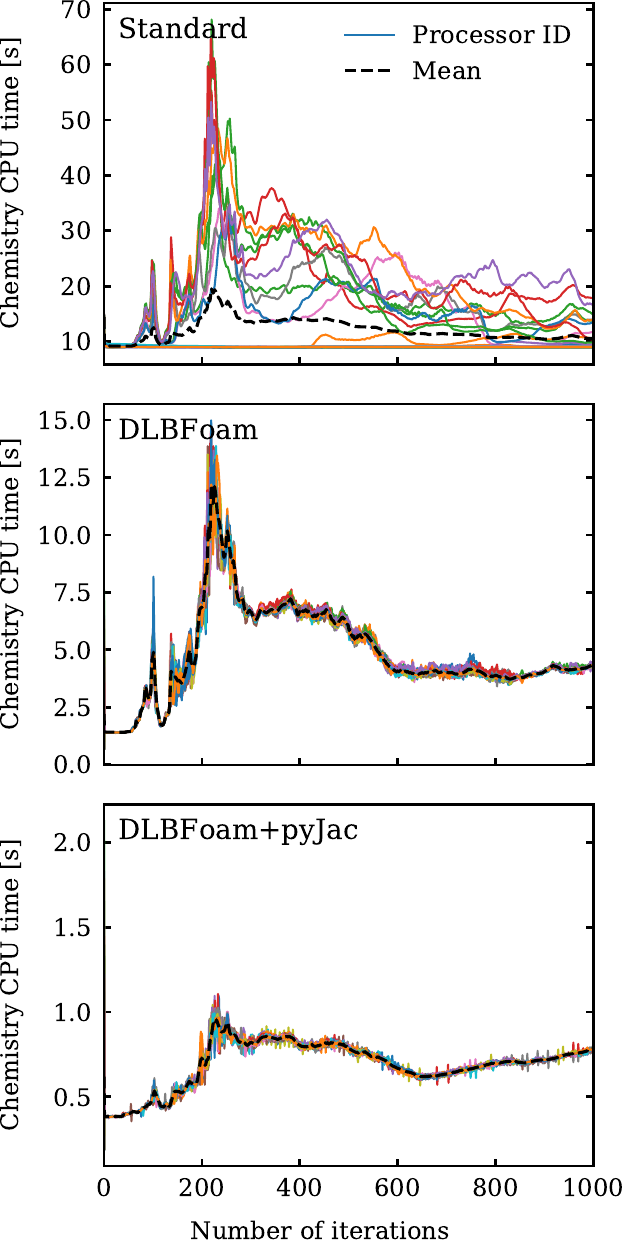}
	    \caption{Chemistry solution CPU time for each process and the mean CPU time of all processes. Data, plotting scripts, and the figure file are available under CC-BY \cite{paperdata}.}
	\label{fig:stats}
    \end{figure}

    Figure \ref{fig:stats} shows the CPU time that each process spends on computing the chemistry in the parallel simulation for 1) Standard, 2) \texttt{DLBFoam}, and 3) \texttt{DLBFoam}+\texttt{pyJac} cases. The arithmetic mean CPU times of the processes are also presented in the figure. It can be seen that for the Standard case, the process-based CPU times are spread widely around the mean value, indicating an imbalanced distribution of the chemistry CPU load. When \texttt{DLBFoam} it utilized, the deviation between the ranks is diminished and a computational balance is established with all the processes having close CPU time values to the mean CPU time. Also, it can be observed that the mean CPU time of the \texttt{DLBFoam} case is also slightly lower than the Standard case due to the utilization of the reference mapping method, as described in the Methodology section. Finally, when \texttt{DLBFoam}+\texttt{pyJac} configuration is used, it can be seen that the chemistry CPU time is decreased even further while still maintaining the balanced load distribution due to the \texttt{DLBFoam}.

\begin{figure}[!h]
    \centering
    \includegraphics[width=67mm]{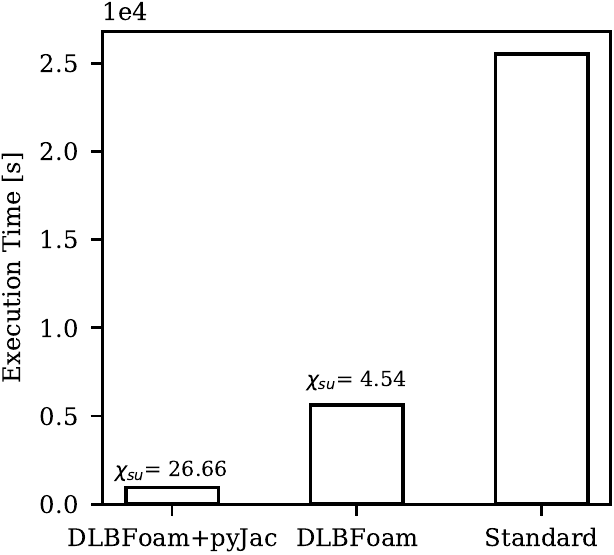}
    \caption{Execution time of the 2D shear layer benchmark case using 3 different approaches. The factor of speed up values for relative to the standard case ($\chi_{su}$) are also provided. Data, plotting scripts, and the figure file are available under CC-BY \cite{paperdata}.}
	\label{fig:shearspeedup}
    \end{figure}

    Figure \ref{fig:shearspeedup} shows the total execution time of each configuration over 1000 CFD iterations. For \texttt{DLBFoam} and \texttt{DLBFoam}+\texttt{pyJac} cases, the relative speed-up with respect to Standard case ($\chi_{su}$) are also presented. It can be seen that while \texttt{DLBFoam} provides a speed up by a factor of 4.54, when ODE solver improvements are also introduced the speed up increases up to a factor of 26.66. Note that the obtained speed up (in particular when \texttt{pyJac} is utilized) compared with standard model depends heavily on the chosen ODE solver tolerances. As demonstrated in Figure \ref{fig:0d}, the performance improvement brought by utilizing \texttt{pyJac}+\texttt{LAPACK} is higher at tighter tolerances. As an example, the same 2D analysis was performed with absolute and relative ODE solver tolerances of  1e$^{-10}$ and  1e$^{-6}$, instead of 1e$^{-8}$ and  1e$^{-5}$, respectively. We noted that while the speed up obtained from \texttt{DLBFoam} remained similar (4.54 to 4.23), the speedup obtained with \texttt{DLBFoam}+\texttt{pyJac} has increased to 101.79 due to enhanced relative performance of \texttt{pyJac}+\texttt{LAPACK} compared with the standard ODE solution approach for tight ODE tolerances (illustration not provided).

\begin{figure}[H]
	    \centering
	    \includegraphics[width=75mm]{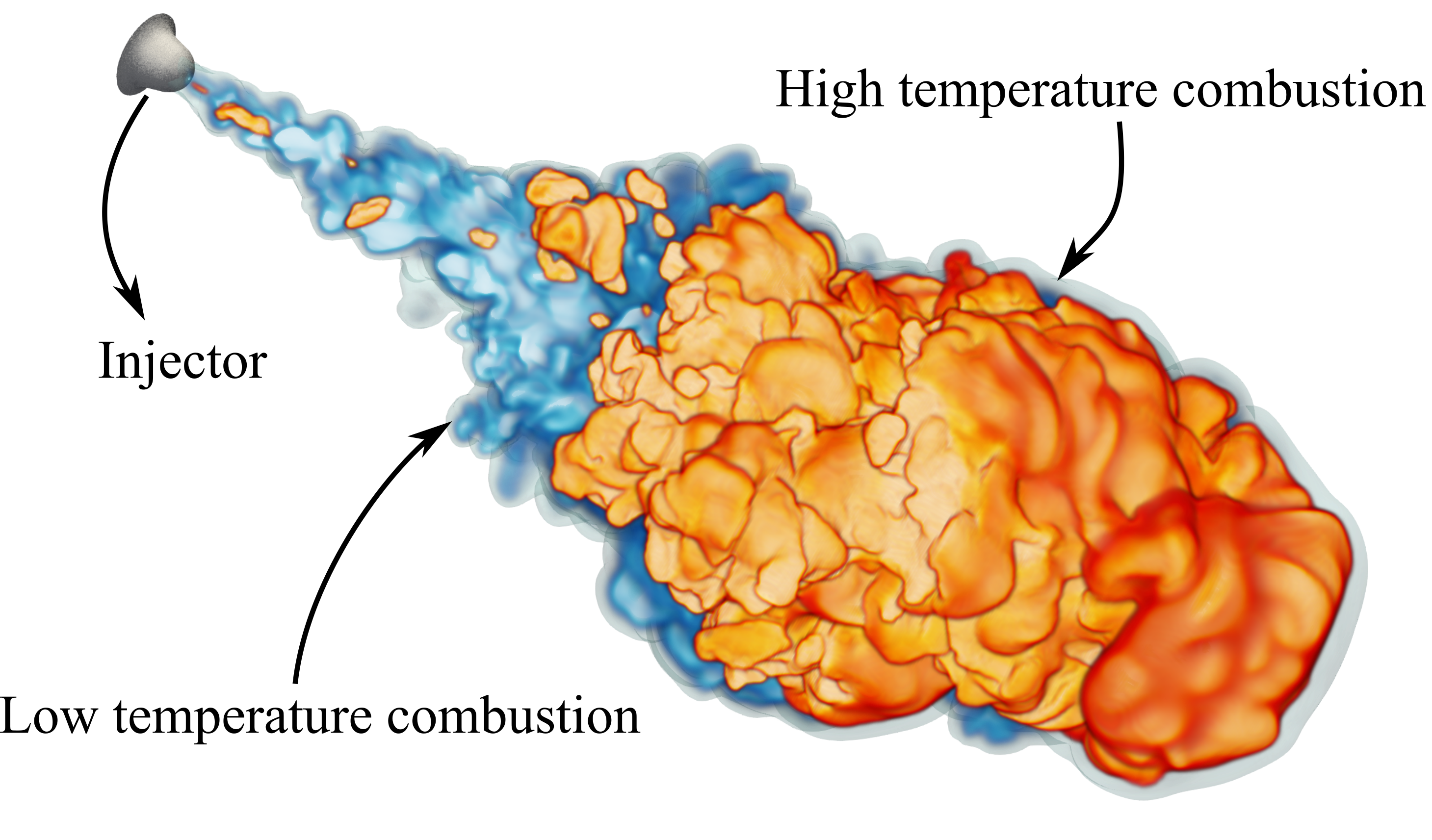}
	    \caption{A volume render of the LES analysis of the ECN Spray A simulation.}
	\label{fig:volrender}
	\end{figure}

    As the final benchmarking case, a 3D large-eddy simulation (LES) of a spray combustion is performed. Similar to our work introduced \texttt{DLBFoam}, the Engine Combustion Network Spray A condition \cite{cite_ECN} is utilized. Figure \ref{fig:volrender} shows a volume render illustrating the spray combustion event. Further details of the numerical configuration can be found in our previous studies \cite{Kahila2019,Tekgul2020,Gadalla2020}. A domain of around 4.5M grid points is decomposed into 256 processors. Absolute and relative ODE solver tolerances of  1e$^{-10}$ and  1e$^{-6}$ are utilized, respectively. It is important to note that due to the very poor performance of the standard OpenFOAM model, a full comparison of the cases from start to finish is not possible in a reasonable computational time. Instead, we simulated each case for 100 CFD iterations after spray ignition ($\approx$\SI{0.3}{\milli\second}) and compared the execution time at that interval.

\begin{figure}[H]
	    \centering
	    \includegraphics[width=67mm]{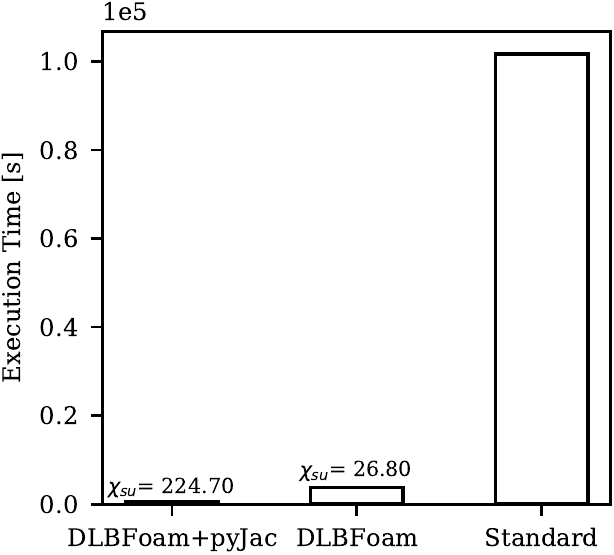}
	    \caption{Execution time of the ECN Spray A simulation for 100 iterations. \texttt{DLBFoam} used together with \texttt{pyJac} provides a speedup of $\chi_{su}$ = 224.70 compared with the standard OpenFOAM chemistry model. Data, plotting scripts, and the figure file are available under CC-BY \cite{paperdata}.}
	\label{fig:spraya}
    \end{figure}

    Figure \ref{fig:spraya} shows the speed up obtained by using \texttt{DLBFoam} and \texttt{DLBFoam}+\texttt{pyJac}, compared with standard OpenFOAM chemistry model. It can be seen that when both \texttt{DLBFoam} and \texttt{pyJac} are utilized, a speed up around two orders of magnitude ($\chi_{su}$=224.70) is obtained compared with the standard model.
    
	\label{section:results}
	\section*{Conclusions}
	In this study, an open-source dynamic load balancing model named \texttt{DLBFoam} \cite{DLBFOAM} with a reference mapping feature is re-introduced to speed up the parallel reactive flow simulations by mitigating the computational imbalance due to chemistry. In addition, an analytical Jacobian formulation (via \texttt{pyJac}) and an improved ODE solver (via \texttt{LAPACK}) are also presented to speed-up the solution of the stiff chemistry ODE problem in each computational cell. The implementation details of the model are described and benchmarking results are given. Utilizing the \texttt{DLBFoam} and \texttt{pyJac} together, we reported a speed-up by a factor of over 200 for large-scale 3D reactive flow simulations. 
	
	Even though OpenFOAM is a very versatile open-source CFD solver for a wide range of different applications, it is unable to compete with its commercial alternatives when it comes to reacting flow simulations due to the issues we described in this paper. Our open-source dynamic load balancing model \texttt{DLBFoam} coupled with the analytical Jacobian routines generated by \texttt{pyJac} and \texttt{LAPACK} ODE routines mitigate most of these issues and make OpenFOAM a viable option for reacting flow simulations.
	\label{section:conclusions}
	\bigskip
	\section*{Acknowledgements}
    The present study has been financially supported by the Academy of Finland (grant number 318024). The computational resources for this study were provided by CSC - Finnish IT Center for Science. The first author has been financially supported by the Merenkulun S\"{a}\"{a}ti\"{o}. 
    \bigskip
    \section*{Supplementary material}

    All the analysis in this paper is performed using the DLBFoam. DLBFoam is open-source and publicly available at \url{https://github.com/Aalto-CFD/DLBFoam}. The repository contains instructions for compilation and tutorials. ODE solver improvements related to \texttt{pyJac} and \texttt{LAPACK} will soon be added to the repository. In addition, the data, plotting scripts, and figure files for a selection of the figures features in this paper are available publicly under a CC-BY license \cite{paperdata}.
    \bigskip
    \bibliography{library.bib} %%User-specified
    \bibliographystyle{elsarticle-num.bst}
	\end{document}